\documentclass[structabstract]{aa}  
\usepackage{graphicx}
\usepackage{txfonts}
\usepackage{natbib}
\usepackage{breqn}
\usepackage{longtable,lscape}
\newcommand{\SiIII}[1]{\mbox{Si\,{\sc iii}~$\lambda${#1}}}
\begin{document}

   \title{Lucky Spectroscopy, an equivalent technique to Lucky Imaging}
   \subtitle{Spatially resolved spectroscopy of massive close visual binaries \\
             using the William Herschel Telescope}


   \author{J. Ma{\'\i}z Apell{\'a}niz\inst{1}
           \and
           R. H. Barb\'a\inst{2}
           \and
           S. Sim\'on-D{\'\i}az\inst{3,4}
           \and
           A. Sota\inst{5}
           \and
           E. Trigueros P\'aez\inst{1,6}
           \and
           J. A. Caballero\inst{1}
           \and
           E. J. Alfaro\inst{5}
          }

   \institute{Centro de Astrobiolog{\'\i}a, CSIC-INTA. Campus ESAC. Camino bajo del castillo s/n. E-28\,692 Vill. de la Ca\~nada. Madrid. Spain. \\
              \email{jmaiz@cab.inta-csic.es} 
         \and
              Departamento de F{\'\i}sica y Astronom{\'\i}a. Universidad de La Serena. Av. Cisternas 1200 Norte. La Serena. Chile. 
         \and
              Instituto de Astrof{\'\i}sica de Canarias. E-38\,200 La Laguna, Tenerife. Spain. 
         \and
              Departamento de Astrof{\'\i}sica. Universidad de La Laguna. E-38\,205 La Laguna, Tenerife. Spain. 
         \and
              Instituto de Astrof{\'\i}sica de Andaluc{\'\i}a-CSIC. Glorieta de la Astronom\'{\i}a s/n. E-18\,008 Granada. Spain. 
         \and
              Departamento de F{\'\i}sica. Ingenier{\'\i}a de Sistemas y Teor{\'\i}a de la Se\~nal. Escuela Polit\'ecnica Superior. Universidad de Alicante. Carretera San Vicente del Raspeig s/n. E-03\,690 San Vicente del Raspeig. Alicante. Spain. 
             }

   \date{Submitted 23 Feb 2018; accepted 06 Apr 2018}

 
  \abstract
  {Many massive stars have nearby companions whose presence hamper their characterization through spectroscopy.}
  {We want to obtain spatially resolved spectroscopy of close massive visual binaries to derive their spectral types.}
  {We obtain a large number of short long-slit spectroscopic exposures of five close binaries under good seeing conditions, select those with the best
   characteristics, extract the spectra using multiple-profile fitting, and combine the results to derive spatially separated spectra.}
  {We demonstrate the usefulness of Lucky Spectroscopy by presenting the spatially resolved spectra of the components of each system, in two cases with separations of 
   only $\sim$0\farcs3. Those are $\delta$~Ori~Aa+Ab (resolved in the optical for the first time) and $\sigma$~Ori~AaAb+B (first time ever resolved). We also spatially
   resolve 15~Mon~AaAb+B, $\zeta$~Ori~AaAb+B (both previously resolved with GOSSS, the Galactic O-Star Spectroscopic Survey), and $\eta$~Ori~AaAb+B, a system with two 
   spectroscopic B+B binaries and a fifth 
   visual component. The systems have in common that they are composed of an inner pair of slow rotators orbited by one or more fast rotators, a characteristic that 
   could have consequences for the theories of massive star formation.}
  {}


   \maketitle
%

\section{Introduction}

$\,\!$ \indent The Galactic O-Star Spectroscopic Survey (GOSSS, \citealt{Maizetal11}) 
is obtaining $R\sim$2500, signal-to-noise ratio S/N~$\gtrsim$~200, blue-violet spectroscopy of all optically accessible O stars in the Galaxy.
To this date, three survey papers (\citealt{Sotaetal11a,Sotaetal14,Maizetal16}, from now on GOSSS~I+II+III) have been published
with a total of 590 O stars\footnote{The GOSSS spectra are being gathered with six facilities: 
the 1.5~m Telescope at the Observatorio de Sierra Nevada (OSN); 
the 2.5~m du Pont Telescope at Las Campanas Observatory (LCO); 
the 3.5~m Telescope at the Observatorio de Calar Alto (CAHA); 
and the 2.0~m Liverpool Telescope (LT),
the 4.2~m William Herschel Telescope (WHT), 
and the 10.4~m Gran Telescopio Canarias (GTC) at the Observatorio del Roque 
de los Muchachos (ORM). Of those, the LT is a recent addition to the mix}. O stars love company and few (if any) 
of them is born completely isolated. As a result, their multiplicity fraction, both visual and spectroscopic is
close to one (see GOSSS~II). This characteristic can be a blessing, allowing for the measurement of masses through
their orbits, but most times is also a curse, as multiplicity can be hard to identify and resolve (spatially or
spectroscopically) and its hidden nature introduces biases when calculating properties such as the initial mass function \citep{Maiz08a} and produces
spectral peculiarities that may be mistaken for unique characteristics of the target.

Lucky Imaging \citep{Lawetal06,Baldetal08,Smitetal09} is a high-spatial resolution passive imaging technique that takes a large number of short exposures, selects the 
ones with the best quality, and combines them to produce a final result with a full width at half maximum (FWHM)
much better than the seeing that would be obtained in a long exposure. In a strict
sense, Lucky Imaging requires that one reaches the diffraction limit and that places restrictions on the wavelength used ($z$ and $i$ bands are preferred over shorter 
wavelengths), the telescope size (small telescopes yield larger diffraction-limited point spread functions (PSFs)
but large telescopes have a lower probability of producing good images, so
a 2-4~m telescope is the best choice), and integration time (which must be at least similar to the atmospheric coherence time determined by turbulence).
In a looser sense, we can still call Lucky Imaging a setup where some of those requirements are not met, yielding a final product with a FWHM improved over that of a 
long exposure but not reaching the diffraction limit. It is that sense that will be used here.

Lucky Spectroscopy is the logical extension of Lucky Imaging to spectroscopy: obtaining a large number of short long-slit spectroscopic exposures under good-seeing
conditions, selecting those with the best characteristics, and combining them to derive spatially resolved spectra of two or more closely separated point sources 
aligned with the slit. In this paper we describe how we have done precisely that for five massive visual multiple systems observed with the WHT as part of GOSSS.

\section{Data and methods}

\begin{table}
\caption{Multiple systems observed in this paper. For each system we give the total number of exposures, the number of exposures used (with two values when the 
system was observed on two nights), and the single exposure time. The systems separation, position angle, and approximate magnitude difference were obtained from 
the Washington Double Star Catalog \citep{Masoetal01} and our own AstraLux data \citep{Maiz10a}}
\label{systems}
\begin{center}
\begin{tabular}{lccccccc}
\hline
System       & Comp.  & $n_{\rm exp}$ & $n_{\rm used}$ & $t_{\rm exp}$ & Sep.      & P.A.     & $\Delta m$ \\
\hline
$\zeta$ Ori  & AaAb+B & 100           &  12            & 0.1 s         & 2\farcs42 & 167\degr & 2.3        \\
$\sigma$ Ori & AaAb+B & 100           &   7            & 1.0 s         & 0\farcs26 &  70\degr & 1.2        \\
15 Mon       & AaAb+B & 110           &  17            & 1.0 s         & 3\farcs00 & 214\degr & 3.1        \\
$\delta$ Ori & Aa+Ab  & 100           &  10/6          & 0.1 s         & 0\farcs30 & 131\degr & 1.3        \\
$\eta$ Ori   & AaAb+B & 100           &  29/22         & 1.0 s         & 1\farcs80 &  77\degr & 1.3        \\
\hline
\end{tabular}
\end{center}
\end{table}

$\,\!$ \indent We obtained spectra for five multiple massive-star systems (Table~\ref{systems}) on the night of 2017 September 7 and repeated the observations on the
following night for two of them ($\delta$~Ori~Aa+Ab and $\eta$~Ori~AaAb+B). The spectra were obtained with the standard GOSSS configuration for the Intermediate
dispersion Spectrograph and Imaging System (ISIS) at the WHT (see Table~1 in GOSSS~III) with three modifications:

\begin{itemize}
 \item A slit width of 0\farcs5 was used instead of the standard 0\farcs9 to increase the spectral resolution to $R\sim 4000$. 
 \item A narrow window of 116 pixels (33\farcs2 at 0\farcs20/pixel) in the spatial direction was used to read the CCD in just 15 s.
 \item Either 100 or 110 short exposures of 0.1 s or 1 s each were obtained for each target.
\end{itemize}

Each of the 100-110 exposures of a given object was extracted individually by fitting a two-component one-dimensional Moffat profile to model the PSF i.e. by 
spatially deconvolving the spectrum at each wavelength. The separation
between the two components was kept fixed from the known astrometry (Table~\ref{systems}) and common values of $\alpha$ and $\beta$ (the Moffat profile parameters) were 
used for both components. The characteristics of the PSF were allowed to change as a function of wavelength but an average profile was generated by subsampling the
pixel scale by a factor of ten using the drizzle method \citep{FrucHook02}, recentering,  and collapsing the result in wavelength. As the trace crosses multiple times 
from one pixel to another and we have almost 4000 valid wavelength points, each subsampled pixel receives $\sim$400 contributions and the spectral information is 
erased in the collapsed result. From the collapsed profile we measured the mean magnitude difference $\Delta m$, the mean seeing (defined as the FWHM of the profile 
for a single component), and a $\chi^2$ value for the fit that measures its goodness.

\begin{figure*}
\centerline{\includegraphics[width=0.50\linewidth]{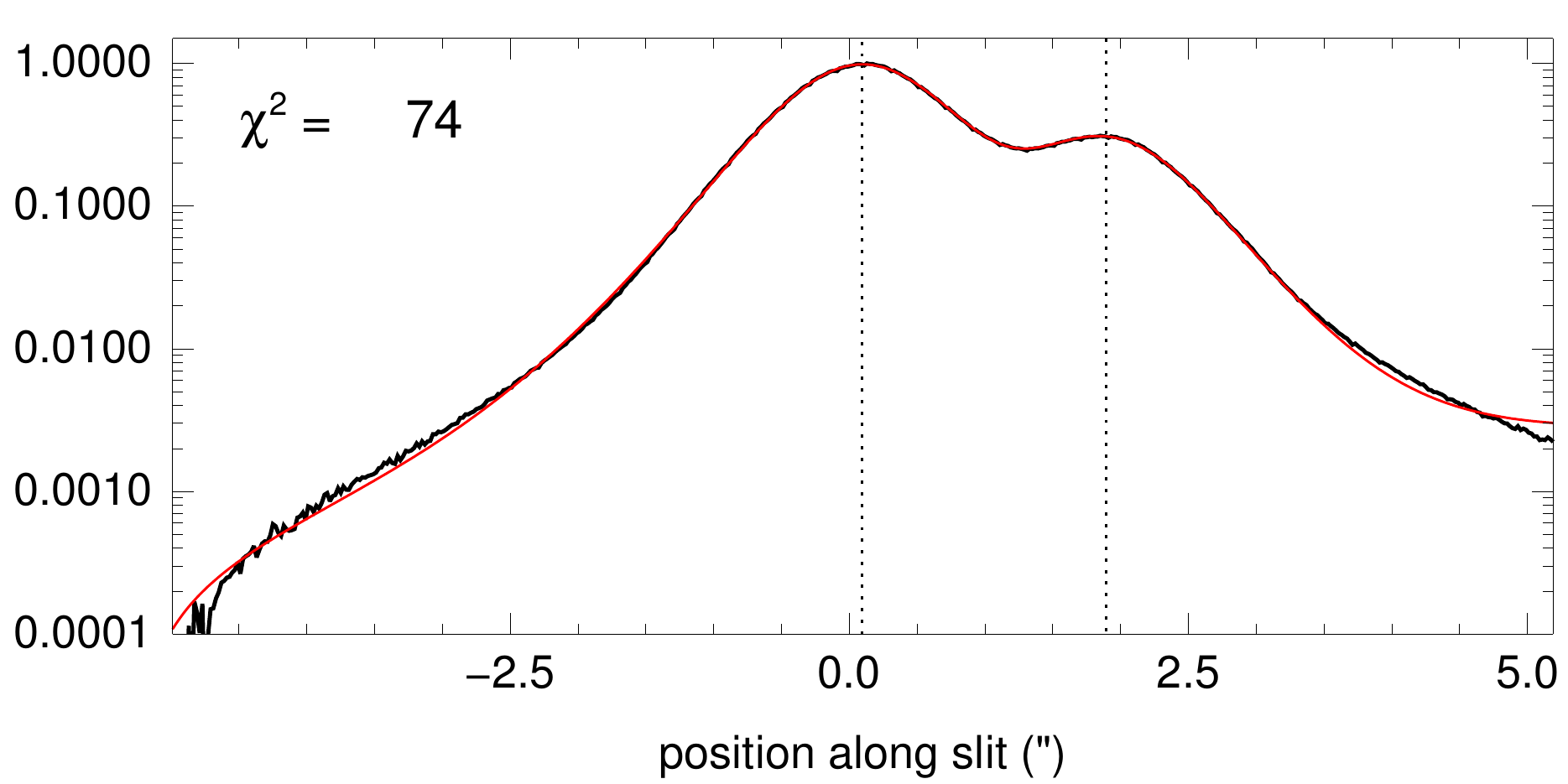}
            \includegraphics[width=0.50\linewidth]{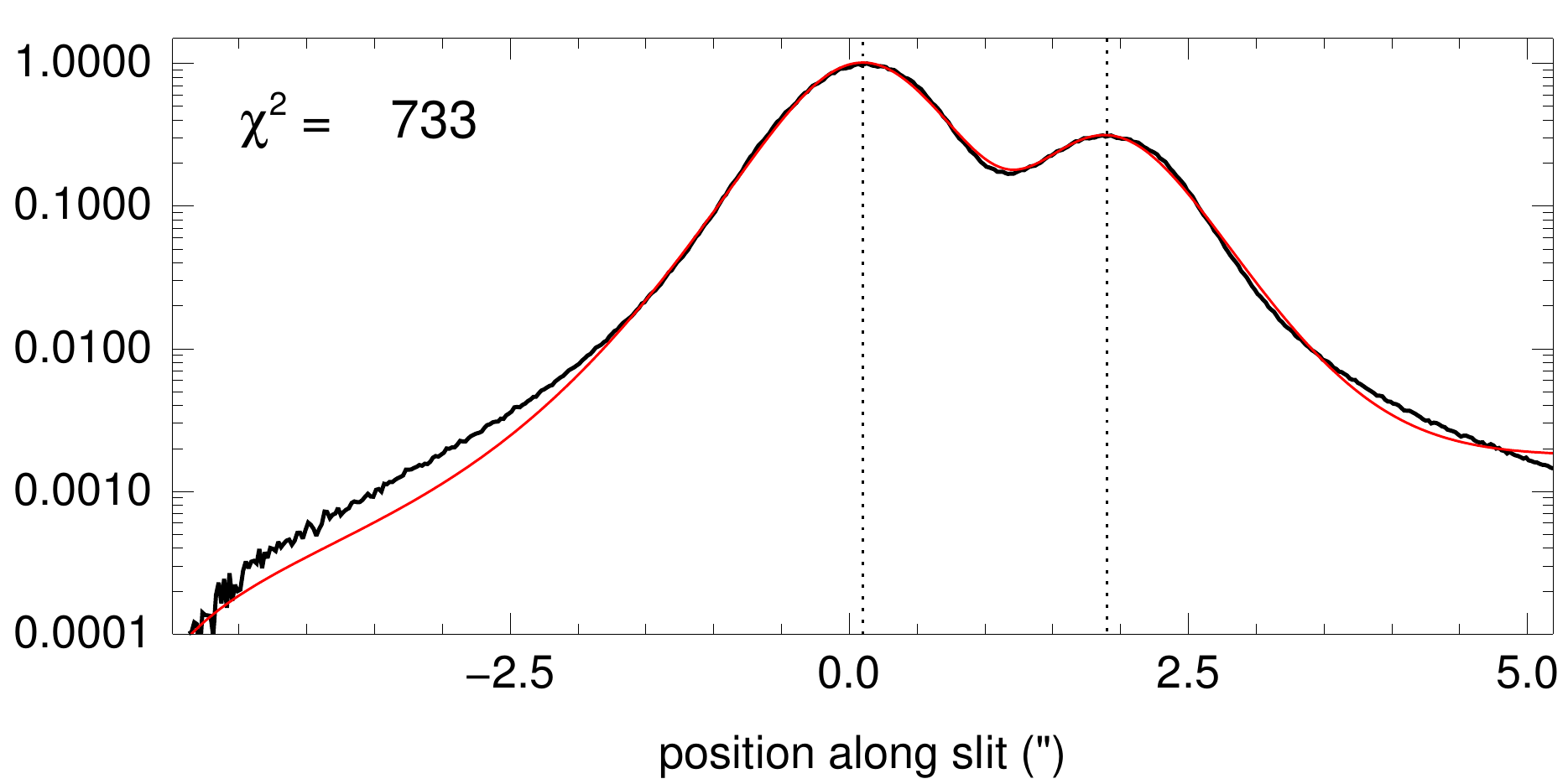}}
\centerline{\includegraphics[width=0.50\linewidth]{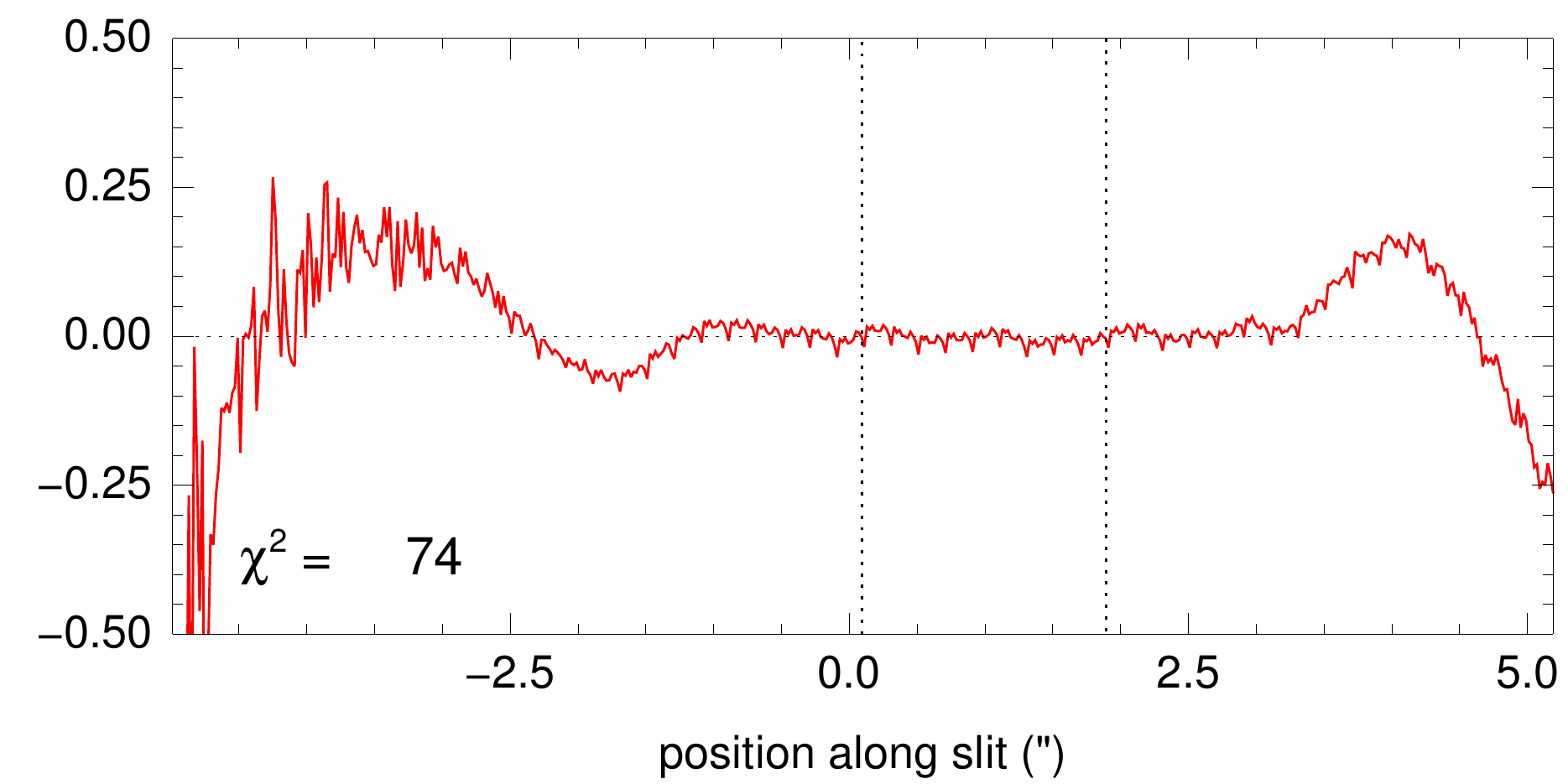}
            \includegraphics[width=0.50\linewidth]{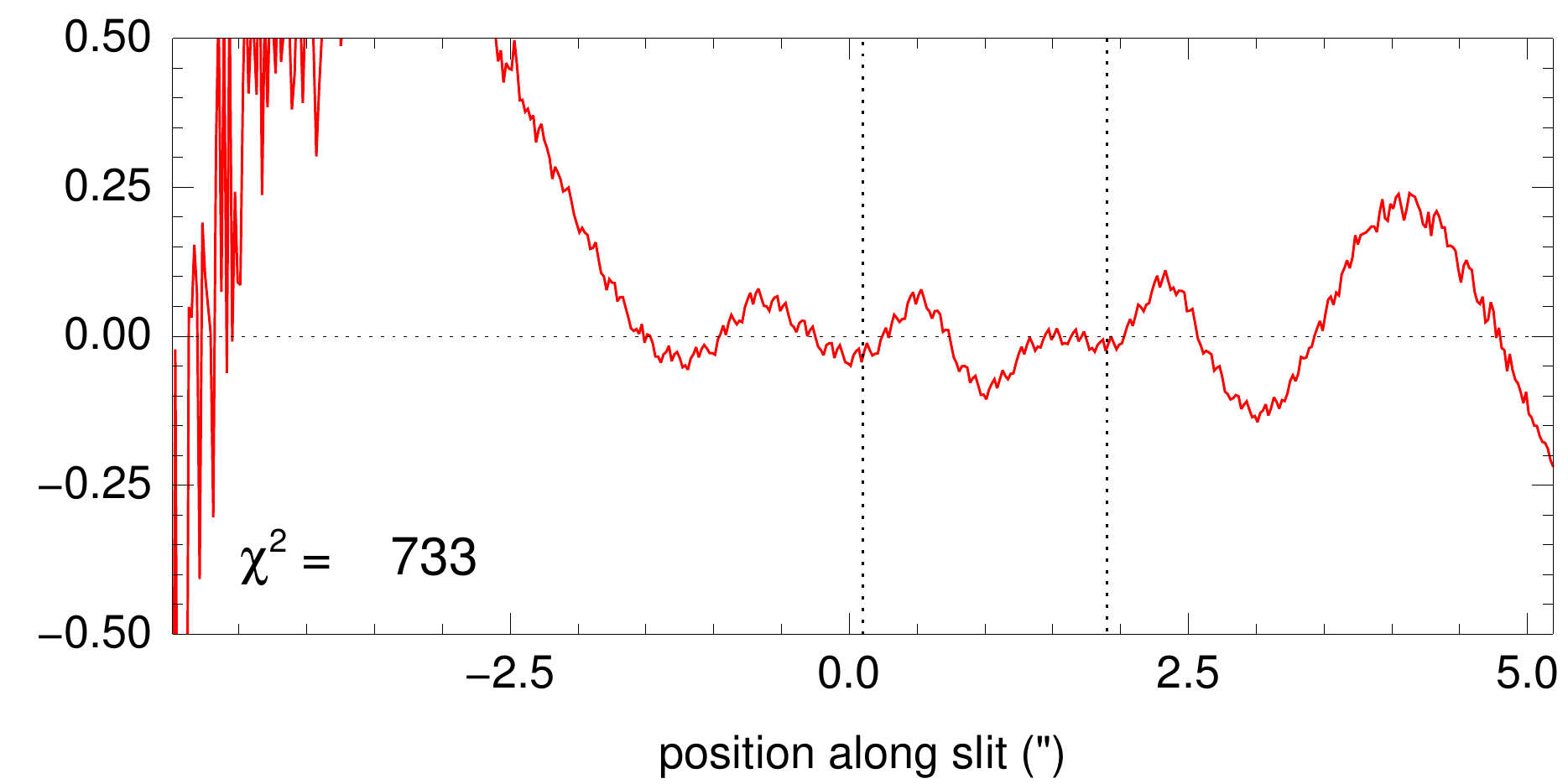}}
\caption{Top: comparison between subsampled collapsed spatial profiles for a selected exposure (left) and a rejected one (right) for the 2017 September 7 observations of 
$\eta$~Ori~AaAb+B. 
{Solid black is used for the observed profile, solid red for the fit, and dotted black for the star positions.} 
Bottom: same as top panels but for the normalized residual i.e. (observed-fit)/fit.
The selected exposure has a more symmetric profile than the rejected one, yielding a much lower $\chi^2$ and a reduced contamination of the primary on the secondary.}
\label{goodbad}
\end{figure*}


Typically for Lucky Imaging, a measurement of the seeing in each frame is obtained and a given fraction (e.g. 1\% or 10\%) of the best frames are selected to produce
the final combined image. Here we follow a similar strategy but using the three parameters above ($\Delta m$, seeing, and $\chi^2$) as conditions instead of just the
seeing. Given that our sample of stars is small and that the number of frames is not too large, we selected the criteria to be applied to each system by
(a) looking at the individual exposures one by one and (b) then selecting a reasonable criterion for each system. Examples of this two-step procedure are shown in
Figs.~\ref{goodbad}.
One important difference in the case of Lucky Imaging is that we combined our selected spectra after extracting and normalizating
them as opposed to selecting the best frames first and then combining them prior to extraction. We were able to do so because of our previous knowledge about the 
expected separation and $\Delta m$ and the high S/N present in the individual exposures compared to a typical Lucky Imaging situation

\begin{figure*}
\centerline{\includegraphics[width=\linewidth]{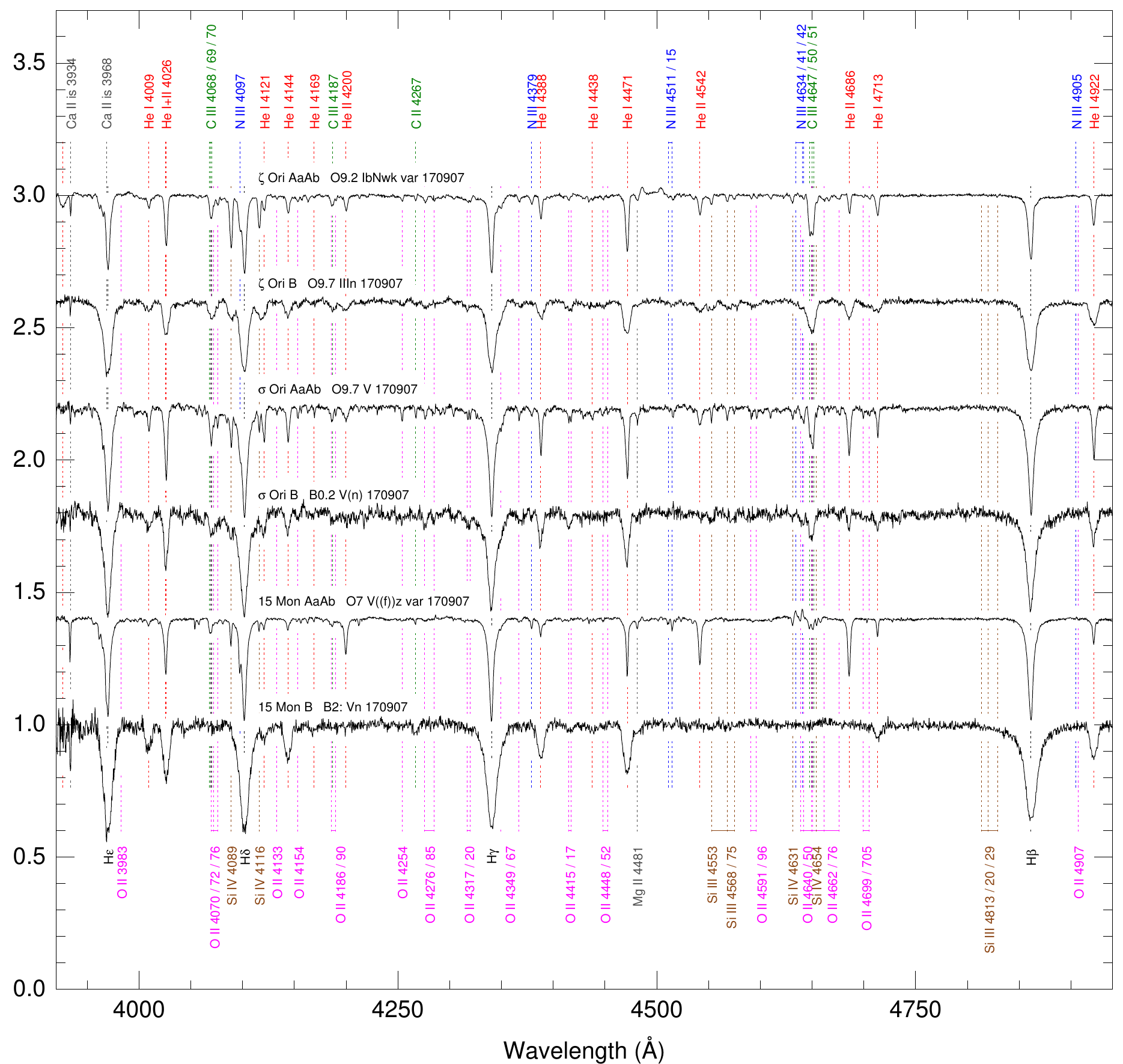}}
\caption{Spectrograms for the three systems observed on a single epoch at the original spectral resolution ($R\sim 4000$) and on the stellar reference frame.
For each spectrogram, the name, spectral type and date (YYMMDD) are shown. Main atomic lines are indicated.}
\label{GOSSS_oneep}
\end{figure*}

\begin{figure*}
\centerline{\includegraphics[width=\linewidth]{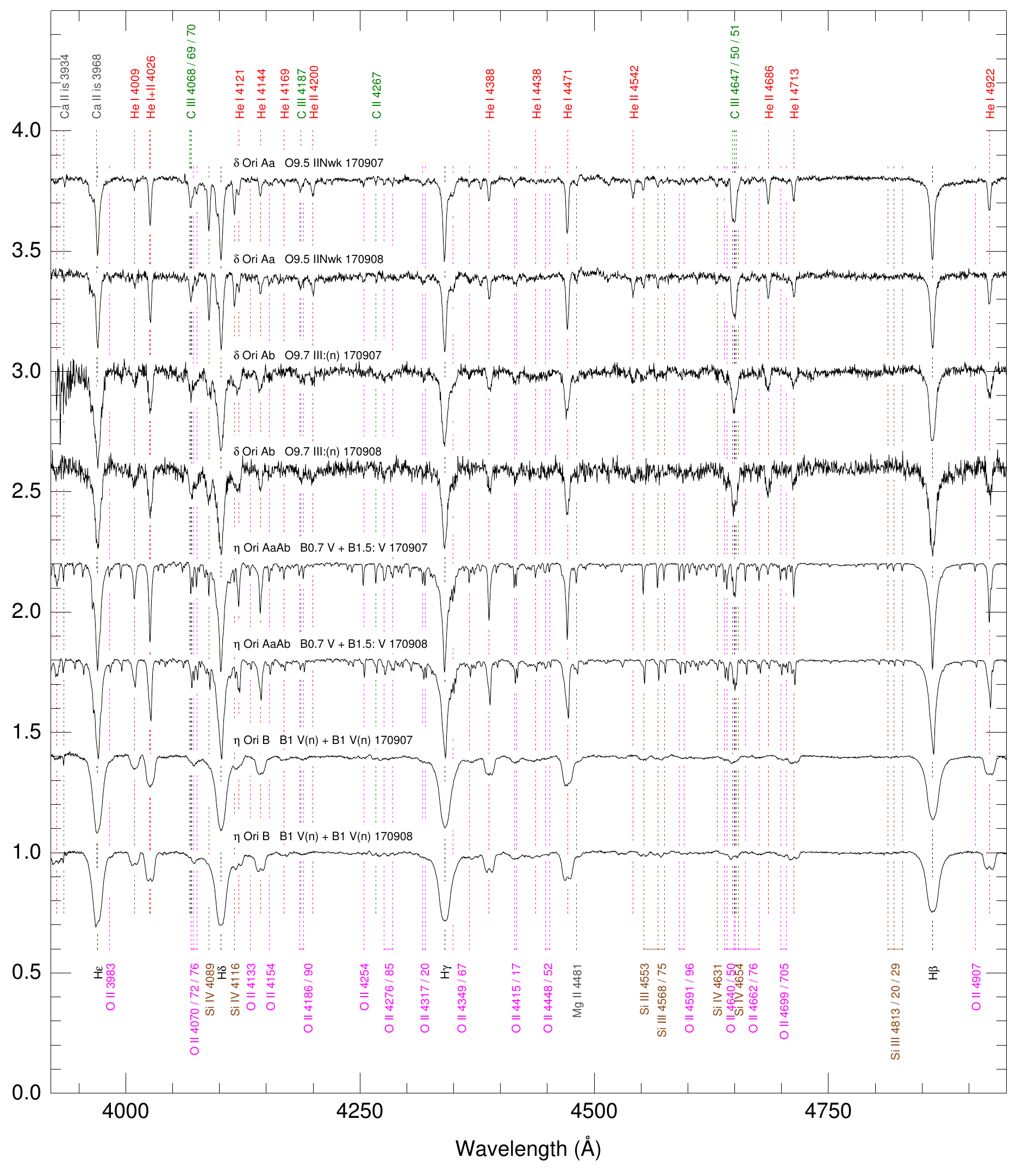}}
\caption{Same as Fig.~\ref{GOSSS_oneep} for the two systems observed on 2017 September 7+8.}
\label{GOSSS_twoep}
\end{figure*}

After obtaining the spectrograms, we performed the spectral classification with MGB \citep{Maizetal12,Maizetal15b} and a new grid of spectroscopic standards 
that includes B-type stars (Ma{\'\i}z Apell\'aniz et al., in preparation). The spectrograms in
Figs.~\ref{GOSSS_oneep}~and~\ref{GOSSS_twoep} and spectral types in Table~\ref{GOSSS_spty} are available from the Galactic O-Star Catalog (GOSC,
\citealt{Maizetal04a,Maizetal17c}). GOSC has been recently moved to {\tt http://gosc.cab.inta-csic.es}
but the old URL ({\tt http://gosc.iaa.es}) will be temporarily kept as a mirror.

\begin{table*}
\caption{New spectral classifications. GOS/GBS stands for Galactic O/B Star. 
The information in this table is also available in electronic form at the GOSC web site
({\tt http://gosc.cab.inta-csic.es}) and at the CDS via anonymous ftp to {\tt cdsarc.u-strasbg.fr} (130.79.128.5) or via 
{\tt http://cdsweb.u-strasbg.fr/cgi-bin/qcat?J/A+A/}.}
\label{GOSSS_spty}
\begin{center}
\begin{tabular}{lcccllll}
\hline
Name              & GOSSS ID               & R.A. (J2000) & Decl. (J2000)  & ST    & LC   & Qual.      & Second.    \\
\hline
$\zeta$ Ori AaAb  & GOS 206.45$-$16.59\_01 & 05:40:45.527 & $-$01:56:33.28 & O9.2  & Ib   & Nwk var    & \ldots     \\
$\zeta$ Ori B     & GOS 206.45$-$16.59\_02 & 05:40:45.571 & $-$01:56:35.59 & O9.7  & III  & n          & \ldots     \\
$\sigma$ Ori AaAb & GOS 206.82$-$17.34\_01 & 05:38:44.765 & $-$02:36:00.25 & O9.7  & V    & \ldots     & \ldots     \\
$\sigma$ Ori B    & GBS 206.82$-$17.34\_02 & 05:38:44.782 & $-$02:36:00.27 & B0.2  & V    & (n)        & \ldots     \\
15 Mon AaAb       & GOS 202.94$+$02.20\_01 & 06:40:58.656 & $+$09:53:44.71 & O7    & V    & ((f))z var & \ldots     \\
15 Mon B          & GBS 202.94$+$02.20\_02 & 06:40:58.546 & $+$09:53:42.23 & B2:   & V    & n          & \ldots     \\
$\delta$ Ori Aa   & GOS 203.86$-$17.74\_01 & 05:32:00.398 & $-$00:17:56.91 & O9.5  & II   & Nwk        & \ldots     \\
$\delta$ Ori Ab   & GOS 203.86$-$17.74\_02 & 05:32:00.414 & $-$00:17:56.91 & O9.7  & III: & (n)        & \ldots     \\
$\eta$ Ori AaAb   & GBS 204.87$-$20.39\_01 & 05:24:28.617 & $-$02:23:49.69 & B0.7  & V    & \ldots     & B1.5: V    \\
$\eta$ Ori B      & GBS 204.87$-$20.39\_02 & 05:24:28.729 & $-$02:23:49.32 & B1    & V    & (n)        & B1 V(n)    \\
\hline
\end{tabular}
\end{center}
\end{table*}


\section{Results}

$\,\!$ \indent We present our results here and we compare them to the previously available information for each system.

\paragraph{$\zeta$~Ori~AaAb+B = Alnitak~AaAb+B = 50~Ori~AaAb+B = HD~37\,742~AB+HD~37\,743 = BD~$-$02~1338~AaAb+B = ALS~14\,793~AB+ALS~16\,893.} 
$\zeta$~Ori is Orion's Belt easternmost star and a a multiple system with three close bright visual components (Aa, Ab, and B) and a fourth distant (53\arcsec) dim one 
(C) that will not be considered here.
Ab is a very close optical companion in a 7-year orbit that is always less than 50 mas away from Aa and is fainter by 2 mag \citep{Hummetal13a}, so the pair cannot 
be spatially resolved in our data and only a combined AaAb spectral type can be derived. The B component is 2\farcs424 away \citep{Maiz10a} and is 2.3 mag 
fainter than the combined light of AaAb\footnote{Due to an uncaught typo, the $\Delta m$ was given as exactly 2.424 in GOSSS~I+II. The actual magnitude difference 
cannot be expressed with such precision unless a specific wavelength is
specified.}. In our data we resolve B from AaAb, something that we already did using conventional techniques in GOSSS~I+II. The 
spectral type we derive for AaAb is identical to that of GOSSS~II, O9.2~IbNwk~var. That of B is slightly different: the spectral subtype changes from O9.5 to O9.7, the 
luminosity class from II-III to III, and the rotation index from (n) to n. The new n rotation index for the B component is consistent with the value of $v \sin i$ of 
350 km/s measured by \citet{Hummetal13a}. The B spectrogram has a very good S/N and no sign of residual contamination from AaAb. In this case, Lucky Spectroscopy 
improves previous results using conventional techniques by reducing the contamination of the primary on the secondary.

\paragraph{$\sigma$~Ori~AaAb+B = 48~Ori~AaAb+B = HD~37\,468~AaAb+B = BD~$-$02~1326~AaAb+B = ALS~8473~AaAb+B.} The $\sigma$~Ori system is a sexdecuple system surrounded by 
its homonymous open cluster \citep{Caba14}. The three brightest stars, Aa, Ab, and B are all contained within 0\farcs3, with
the rest of the components located at significantly larger separations. The B component is in a nearly circular orbit around AaAb with a period of 159.90~a and a 
current separation of 0\farcs26 \citep{Schaetal16}. The Aa and Ab components form a binary whose existence was not confirmed through spectroscopic means until very 
recently \citep{SimDetal11b}. The Aa+Ab system has an eccentric orbit with a period of 143.198~d \citep{SimDetal15a} that has been resolved with interferometry 
\citep{Schaetal16}. We have been able to spatially resolve B from AaAb, which to our knowledge had never been done with spectrosccopy before, making it the system
in this paper resolved with the 
{smallest} 
separation. The $\sigma$~Ori~AaAb spectrum in Fig.~\ref{GOSSS_oneep} has a very good S/N and yields a combined spectral type of 
O9.7~V. The spectral subtype is the same as the one in GOSSS~II, but the luminosity class there is III instead of V. As discussed in GOSSS~III, we expected that
previous luminosity classification to be the result of the contamination from B, given the young age of the system, and our results here confirm that
(see also \citealt{SimDetal15b}, where they catch the spectroscopic binary with a large velocity difference and are able to provide separate information for Aa and Ab; 
it is our intention to reobserve the system with Lucky Spectroscopy at such a moment to provide separate spectral types for Aa and Ab).
The $\sigma$~Ori~B spectrum is noisier, as expected for the dimmer companion of a spatially
deconvolved system, but there are no strong signs of contamination from AaAb (e.g. compare \SiIII{4552} for the two of them). We derive a spectral type of B0.2~V(n),
which is consistent with the expected result from \citet{SimDetal15b}, both in terms of spectral type and rotational velocity.

\paragraph{15~Mon~AaAb+B = S~Mon~AaAb+B = HD~47\,839~AaAb+B = BD~$+$10~1220~AaAb+B = ALS~9090~AaAb+B.} 15~Mon is similar to $\sigma$~Ori in that it is a system with 
many components inside a small cluster, NGC~2264 \citep{Walk56,Rebuetal02}.
The three brightest stars, Aa, Ab, and B are all contained within 3\arcsec, with the rest of the components located at significantly larger 
separations. The inner pair Aa+Ab has a $\Delta m$ of 1.2 mag, is currently separated by 0\farcs1, and its orbit is rather uncertain, with a high eccentricity
and a period measured in 
decades \citep{Giesetal97,Cvetetal10}. The B component is located 3\arcsec\ away from AaAb and the $\Delta m$ is 3.1 magnitudes, the largest difference in our sample.
Our data cannot resolve Aa and Ab but we obtain separate spectra for AaAb and B, something that we did previously using conventional techniques in GOSSS~I+II. The AaAb
spectrogram has an excellent S/N and its spectral type is the same as our previous GOSSS one, O7~V((f))z~var. The variability may be due to the profile differences
caused by the orbit of Ab around Aa. For the B component, we had previously obtained different spectral classifications as a late O or an early B star due to the
limitations of the conventional techniques. With Lucky Spectroscopy we establish it is an early B star with a classification of B2:~Vn, which is consistent with the
observed $\Delta m$. The uncertainty in the spectral subtype is due to the relatively poor S/N of the deconvolved spectrum. There are no signs of 
contamination from AaAb. as the two spectra have very different rotational velocities: B is a fast rotator while the combined AaAb spectrum shows very narrow lines.

\paragraph{$\delta$~Ori~Aa+Ab = Mintaka~Aa+Ab = 34~Ori~Aa+Ab = HD~36\,486~A+B = BD~$-$00~983~A+B = ALS~14\,779~A+B.} $\delta$~Ori is Orion's Belt westernmost star 
and a multiple system at the center of the Mintaka cluster \citep{CabaSola08}. It has two close bright visual components (Aa and Ab) and two distant dim ones (B and C) 
that will not be considered here. Aa is itself composed of two spectroscopic components (Aa1+Aa2) in an eclipsing orbit with a 5.7325~d period, while Ab is two magnitudes 
fainter than the two Aa components and located 0\farcs267 away in 1993 and moving away from the primary \citep{Hart04,Harvetal02}. We observed the system on two
consecutive nights and in both cases we were able to spatially resolve Ab from Aa, this being one of the two systems in this paper ($\sigma$~Ori is the other one) 
where the power of Lucky Spectroscopy becomes more apparent. The two spectra for Aa yield the same spectral type as for the combined Aa+Ab value
from GOSSS I, O9.5~IINwk, but the lines are slightly narrower. When analyzing the data for GOSSS I the combined spectrum was close to being (n), something that does
not happen for the spatially resolved spectra (this has the additional advantage of giving us a more pure spectrum for a classical O9.5~II classification standard).
There are very small variations between the two epochs for Aa, likely due to the motion of Aa1 and Aa2 but the signature of the latter in the combined 
spectrum is very weak \citep{Shenetal15}. The two epochs for Ab yield the same spectral type, O9.7~III:(n), and in both cases there are indications of only a slight 
residual contamination from Aa, with the second epoch being noisier. The (n) qualifier indicates that Ab is a fast rotator, as previously noted by \citet{Harvetal02} 
and \citet{Shenetal15}. 
The differences between the spectral types of Aa (Aa1+Aa2) and Ab are consistent with the $T_{\rm eff}$ and $\log g$ differences measured by \citet{Shenetal15} using
spectral disentangling (as opposed to the spatial deconvolution used here). \citet{Richetal15} also spatially resolved Aa and Ab in the UV using HST/STIS and obtained very 
similar values of $T_{\rm eff}$ around 31~kK for Aa1 and Ab, with error bars close to 2000~K, which is also consistent with our spectral classifications. That paper
using UV data is the only one we have found where $\delta$~Ori~Aa+Ab was spatially deconvolved, making our result the first time it has been done in the optical.

\paragraph{$\eta$~Ori~AaAb+B = 28~Ori~AaAb+B = HD~35\,411~AaAb+B = BD~$-$02~1235~AaAb+B = ALS~14\,775~AaAb+B.} $\eta$~Ori is a multiple system with no associated
open cluster \citep{CabaSola08} and with three
close bright visual components (Aa, Ab and B) and a distant dim one (C) that will not be considered here. The Aa+Ab pair has a separation of less than 0\farcs1 and a 
$\Delta m$ of 1.5 mag in a 9.442~a orbit \citep{Baleetal99}, making it too close to be resolved in our data. On the other hand, B is separated from the AaAb system by 
1\farcs8 with a $\Delta m$ of 1.3 mag, making it a relatively easy target for Lucky Spectroscopy ($\Delta m$ similar to $\sigma$~Ori~AaAb+B or $\delta$~Ori~Aa+Ab but at a 
larger separation). $\eta$~Ori~AaAb+B differs from the other four systems in this paper in not including an O star (it had not appeared in a GOSSS paper before) and 
in having a complex photometric curve. One photometric period is 7.989255~d long and is associated with the eclipsing binary Aa (unresolved components Aa1+Aa2,
\citealt{WaelLamp88,deMeetal96}) while a
second, shorter one, has an unclear origin. It has been proposed that it originates in B being another eclipsing binary with a 0.864~d period \citep{WaelLamp88,Leeetal93b},
but it has not been proved conclusively, as other alternatives are possible and no spectrum of B separated from AaAb had been previously published. In our Aa+Ab
spectrograms two narrow-lined B-type components are seen, with the weak component moving redwards with respect to the strong one in the 1.023~d elapsed between the 
two epochs.  The derived spectral types are B0.7~V and B1.5:~V, and both move in opposite directions between the two epochs, indicating that they correspond to the two
components in Aa, the 7.989255~d spectroscopic binary. The Ab component leaves no apparent signal in the spectrum, even though it could easily do it with a $\Delta m$ 
of 1.5 mag. One possibility is that it is a fast rotator (see \citealt{SimDetal15a} for the discussion of this possibility for $\sigma$~Ori~B, which we have confirmed 
here). The B spectrograms show two nearly identical spectroscopic components, two fast rotators with spectral types B1~V(n). The two epochs are relatively similar but
the velocity spacing between the two components is larger in the second one ($440\pm20$ km/s) than in the first one ($380\pm20$ km/s). If the period is 0.864~d, the
epoch difference would correspond to 1.18 periods i.e. the stars would have completed a little over one orbit, with the first epoch corresponding to a point just before
quadrature and the second epoch to a point just after quadrature. With such a short period and large velocity differencess the two stars must be in a contact or 
overcontact configuration with a large inclination, making eclipses unavoidable and thus confirming the origin of the \citet{Leeetal93b} light curve. More epochs are
needed to study this system in depth. It could become a future gravitational wave source if both components explode as SNe independently or a
stellar merger if the fusion takes place before that point. In summary, $\eta$~Ori is a complex quintuple system of which four (the two spectroscopic components in
Aa and the two in B) have confirmed early-B spectral types and the fifth (Ab) is also likely to fall in the same category.

\section{Summary and future work}

$\,\!$ \indent We have used Lucky Spectroscopy to obtain spatially resolved spectra of five close massive multiple systems, two of them for the first time (one in the 
optical and the other in any range), and in the process established the validity of the technique. In four cases ($\zeta$~Ori, $\sigma$~Ori, 15~Mon, and 
$\delta$~Ori) we have the same configuration: a hierarchical triple system with an inner pair composed of two slow rotators and an outer component that rotates
significantly faster than the other two. For the fifth system ($\eta$~Ori), there is also an inner pair made up of two slow rotators but with an intermediate orbit 
(that of Ab) where the large rotational velocity is only suspected and an outer system that is itself a pair of fast rotators. This pattern may be related to the
formation mechanisms of massive stars, indicating that the less massive outer object tends to either be the result of mergers or in some other way absorb additional 
rotational angular momentum. Finally, if we combine the results of GOSSS~I+II+III with the recent ones of Ma{\'\i}z Apell\'aniz (2018, submitted to A\&A) we are left 
with the following statistics for the project: 594 O-type systems, 24 of early type other than O (B, A supergiants, sdO, PNN), and 11 of later types.

{Having demonstrated the validity of Lucky Spectroscopy, we plan to use it again with a larger sample to test its limits. As with most techniques that extract 
information from two closely separated point sources, there is likely a region in the separation-$\Delta m$ plane where the technique works well and another one where it does
not. The results here indicate that Lucky Spectroscopy with our setup is capable of separating stars 0\farcs3 apart if $\Delta m$ is small and stars 3\arcsec\ apart with
a $\Delta m \approx 3$ but more observations are needed to map the feasibility boundary. Another line of work we plan to explore is the use of an electron-multiplying
CCD as a detector to reach fainter objects.} 

\begin{acknowledgements}
As we were about to submit this paper, we received the sad news of the death of Nolan R. Walborn after a long battle with cancer. As he was the inspiration for GOSSS,
we would like to dedicate this paper to him. May he rest in peace. This work has made use of the Washington Double Star Catalog \citep{Masoetal01}.
The authors would like to thank the personnel of the Isaac Newton Group of telescopes for their support throughout the years, with a special mention to Cecilia
Fari{\~n}a for her help with the setup for Lucky Spectroscopy. The first author is amazed that the instrument he used for his PhD thesis is still working 
one quarter of a century later and is not yet surpassed in some of its characteristics. We acknowledge support from the Spanish Government Ministerio de Econom{\'\i}a 
y Competitividad (MINECO) through grants AYA2016-75\,931-C2-1/2-P (J.M.A., A.S., E.T.P., and E.J.A), AYA2015-68\,012-C2-1/2-P (S.S.-D. and E.T.P.) and 
AYA2016-79\,425-C3-2-P (J.A.C.). R.H.B. acknowledges support from the ESAC Faculty Council Visitor Program.
\end{acknowledgements}

\vfill

\eject

\bibliographystyle{aa}
\bibliography{general}

\end{document}